\documentstyle[11pt,newpasp_rev,twoside,epsfig]{article}
\begin{document}
\title{Orbital structure of triaxial galaxies} 
 \author{Glenn van de Ven, Ellen Verolme, Michele Cappellari, Tim de
 Zeeuw} 
 \affil{Sterrewacht Leiden, Postbus 9513, 2300 RA Leiden, The
 Netherlands}
 \vspace{-10pt}
 \author{\small glenn@strw.leidenuniv.nl}

\begin{abstract}
  We have developed a method to construct realistic triaxial dynamical
  models for elliptical galaxies, allowing us to derive best-fitting
  parameters, such as the mass-to-light ratio and the black hole mass,
  and to study the orbital structure.  We use triaxial theoretical
  Abel models to investigate the robustness of the method.
\end{abstract}


\section{Triaxial dynamical models}

Many elliptical galaxies show significant signatures of triaxiality
(e.g. de Zeeuw et al. 2002).  Therefore, we have extended
Schwarzschild's orbit superposition method to construct realistic
triaxial dynamical models, which fit the observed surface brightness,
as well as (two-dimensional) kinematical measurements of elliptical
galaxies (Verolme et al. 2003). This fully numerical method is,
however, too computationally expensive to do a full search over the
model parameters, such as mass-to-light ratio, black hole mass,
viewing direction and intrinsic shape. Approximating the potential
by one of St\"ackel form, we can construct velocity and velocity
dispersion fields using the analytical solution of the continuity
equation and the three Jeans equations (Statler 1994, van de Ven et
al. 2003), and compare them with observations to constrain the large
parameter range.  Within this reduced parameter space, we can then
apply the extended Schwarzschild method using the true potential, to
find the true best-fitting triaxial model.  Schwarzschild's method not
only provides the best-fitting parameters, but also results in an
orbital weight distribution, which after appropriate smoothing allows
us to study the orbital structure of the observed galaxy.


\section{Triaxial Abel models}

To investigate the robustness of the derived internal structure, we
are applying our method to theoretical models with known distribution
function (DF).  We use Abel models (Dejonghe \& Laurent 1991; Mathieu
\& Dejonghe 1999), for which the potential is assumed to be of
St\"ackel form and the DF to be a function of a single parameter
$F(E,I_2,I_3)=F(S)$, with $S=E+wI_2+uI_3$ a linear combination of the
explicitly known integrals of motion. The density and higher velocity
moments can be calculated efficiently. Besides this analytical
simplicity, the Abel models have, with the choice of a three-integral
DF, enough freedom to incorporate many of the observed triaxial
features (Figure 1).



\begin{figure}
\vspace{-10pt}
\begin{center}
\epsfig{figure=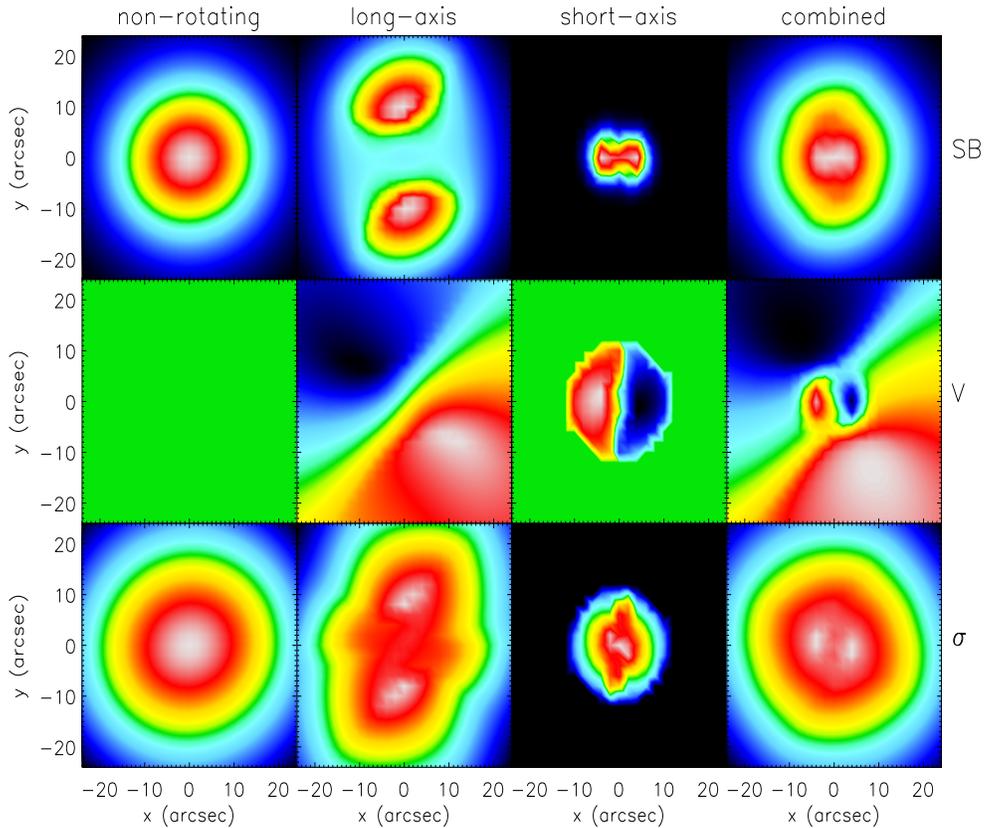,height=11cm,width=13cm}
\vspace{-10pt}
\caption{Kinematics of a triaxial Abel model.
  From top to bottom: the surface brightness (SB), the mean
  velocity ($V$) and velocity dispersion ($\sigma$).  
  From left to right: non-rotating, (outer and inner) long-axis and
  short-axis rotating components, and a (luminosity weighted) combination.  
  The non-rotating components have zero mean velocity. 
  The short-axis component is limited due to the finite extent of Abel
  models.
  The kinematically decoupled core in the combined model is a typical
  signature of triaxiality observed in e.g. NGC4365 
  (Davies et al. 2001).
}
\end{center}
\vspace{-20pt}
\end{figure}

\end{document}